# Mechanical behavior of composite double wall nanotubes from carbon and phosphorous

Kun Cai [1, 2], Jing Wan [1], Likui Yang [1], Ning Wei [1*]

1 College of Water Resources and Architectural Engineering, Northwest A&F University, Yangling 712100, China
2 Research School of Engineering, the Australian National University, ACT, 2601, Australia
* Corresponding author's email address: Nwei@nwsuaf.edu.cn (Ning Wei)

Abstract
Black phosphorus is not stable when it is exposed to air. When covered or terminated by single layer carbon atoms, such as graphene/carbon nanotube, it is more strongly protected in the rapid degradation than the bare black phosphorus. What's more, due to weak interaction between phosphorus atoms in black phosphorene, the nanotube obtained by curling single-layer black phosphorus is not as stable as a carbon nanotube (CNT) at finite temperature. In present work, we recommend a new 1D material, i.e., composite double-wall nanotubes (DWNTs) from a black phosphorus nanotube (BPNT) with a CNT. The dynamic response of the composite DWNTs is simulated using molecular dynamics approach. The effects of such factors as temperature, slenderness and configurations of DWNTs are discussed. Comparing with a single-wall BPNT, the composite DWNTs under uniaxial compression shows some peculiar properties. When the BPNT is embedded in a CNT, the system will not collapse rapidly even if the BPNT has been damaged seriously.

## 1 Introduction

Low-dimensional materials, including such 2D materials as graphene,[1,2] boron nitride,[3,4] MoS2[5,6] and black phosphorene),[7-9] and 1D materials like carbon nanotube (CNT),[10-12] nanoscrolls[13] and nanowires,[14] attracts so much attention in recent years, due to their excellent physical properties. For example, graphene and carbon nanotubes have super high modulus, flexibility and inter-shell/layer superlubrication,[2,15-17] which are essential for a dynamicnanodevices (e.g., nanomotor,[12,18,19] nanobearing,[16,20,21] nano oscillator,[22,23] etc.). For few-layer MoS2/black phosphorene, their perfect electronic properties imply potential design of new semiconductor-based devices. Especially, for a 10-nm black phosphorus sheet, its electron mobility can be 1000cm$^2$/V/s.[7] Using the first principle calculations, Liu et al.[24] found that single-layer black phosphorene has direct gap and the gap depends on layers of black phosphorene. And the dependency was also verified by experiments.[25] Rodin et al.[26] found that the band gap structure of single-layer black phosphorene can be adjusted by introducing in-plane strain. Qiao et al.[27] discussed the anisotropy of electron mobility of single-layer black phosphorene. Fei and Yang[28] found that the electron mobility of black phosphorus is sensitive to in-plane strain. Besides, the optical and photovoltaic properties of black phosphorus are also charming.[29-32] The mechanical properties of black phosphorus have been studied by lots of researchers[33-37] and similar conclusions were obtained, e.g., the in-plane modulus and strength of black phosphorus are far lower than those of graphene. Hence, before using black phosphorus in a nanodevice, one should check its strength and stability firstly.

However, black phosphorus is not stable and reaction with oxygen and water when it is under ambient environment.

Kim et al.[38] showed that the multiple-coverage by graphene strongly enhanced the stability of black phosphorus. Although CNTs are obtained by synthetic approaches, geometrically, people image that a CNT can be formed by curling a graphene ribbon along specific direction. Similarly, people now consider that a nanotube can also be obtained by curling a single-layer black phosphorene (see Figure 1c) even if the tube is still not discovered in laboratory. For black phosphorus nanotubes (BPNTs), their mechanical and electric properties have been estimated recently. For example, Guo et al.[39] provided their prediction on the band gap structures for different BPNTs. Based on density functional theory, Hu et al.[40] found that the band gap depends on the diameter of BPNT, e.g., the gap is lower for a BPNT with larger diameter. Recently, Cai, Wan, et al.[41] considered that higher curvature of a BPNT leads to greater variation of bond lengths and bond angles on the tube. At high temperature, thermal vibration of the phosphorus atoms will enhance such variation. It implies that the tube is not stable at high temperature. Their results provide the relationship between the minimal diameter of a BPNT and temperature. Cai, Wan, et al.[42] discussed the strength and stability of a BPNT under uniaxial compression. Their numerical results demonstrate that the BPNT has low stability under compression and the tube is broken rapidly after buckling. The reason is that the $sp^3$-$sp^3$ phosphorus-phosphorus (P-P) bond has much lower strength than that of the $sp^2$-$sp^2$ carbon-carbon (C-C) bond. Compared with a 2D material, 1D material plays peculiar role in design of NEMS. When we need the special electronic property of BPNT as a component in a dynamic nanodevice, its strength should be improved. Considering the requirement, CNT is a perfect candidate material to provide the BPNT component such protection. In present study, we use molecular dynamics simulation approach to reveal the mechanical properties of such composite nanotubes.

## 2 Models and methodology
### 2.1 Model description

In the present study, the double wall nanotubes (Figure 1d) are made from a single carbon nanotube (CNT) and a BPNT. Three major factors, i.e., temperature of environment, slenderness and configurations of DWNTs, are under consideration as the DWNTs are under dynamic compression. Configurations of DWNTs depends on the selection of the inner and outer tubes. Five schemes of DWNTs are considered in simulation. For simplicity in description, we simplify "armchair CNT" as "C-A", and similarly, "zigzag CNT" as "C-Z", "armchair BPNT" as "P-A" and "zigzag BPNT" as "P-Z", respectively. The geometric parameters of the five schemes are listed in Table 1. Temperature of environment is essential for a nanosystem running in a canonical NVT ensemble. Four temperatures are considered, e.g., T=8, 150, 300 and 400K, respectively, when discussing the dynamics response of the composite DWNTs with $\alpha$=8 in S-1 (in Table 1). Slenderness, which reflects the compliance of DWNTs, is discussed using the variation of α, the ratio between the effective length (Figure 1e) of tube and the average diameter of BPNT, e.g., $d_\mathrm{In}$ in Figure 1a. The ratio are set to be 6, 8, 10 and 12, respectively, when calculating the dynamic response of DWNTs in S-1 and S-2 (in Table 1) at 300K. The effective lengths of the BPNT are 17.59 nm ($\alpha$=6), 23.45 nm ($\alpha$=8), 29.31 nm ($\alpha$=10) and 35.17 nm ($\alpha$=12), respectively.

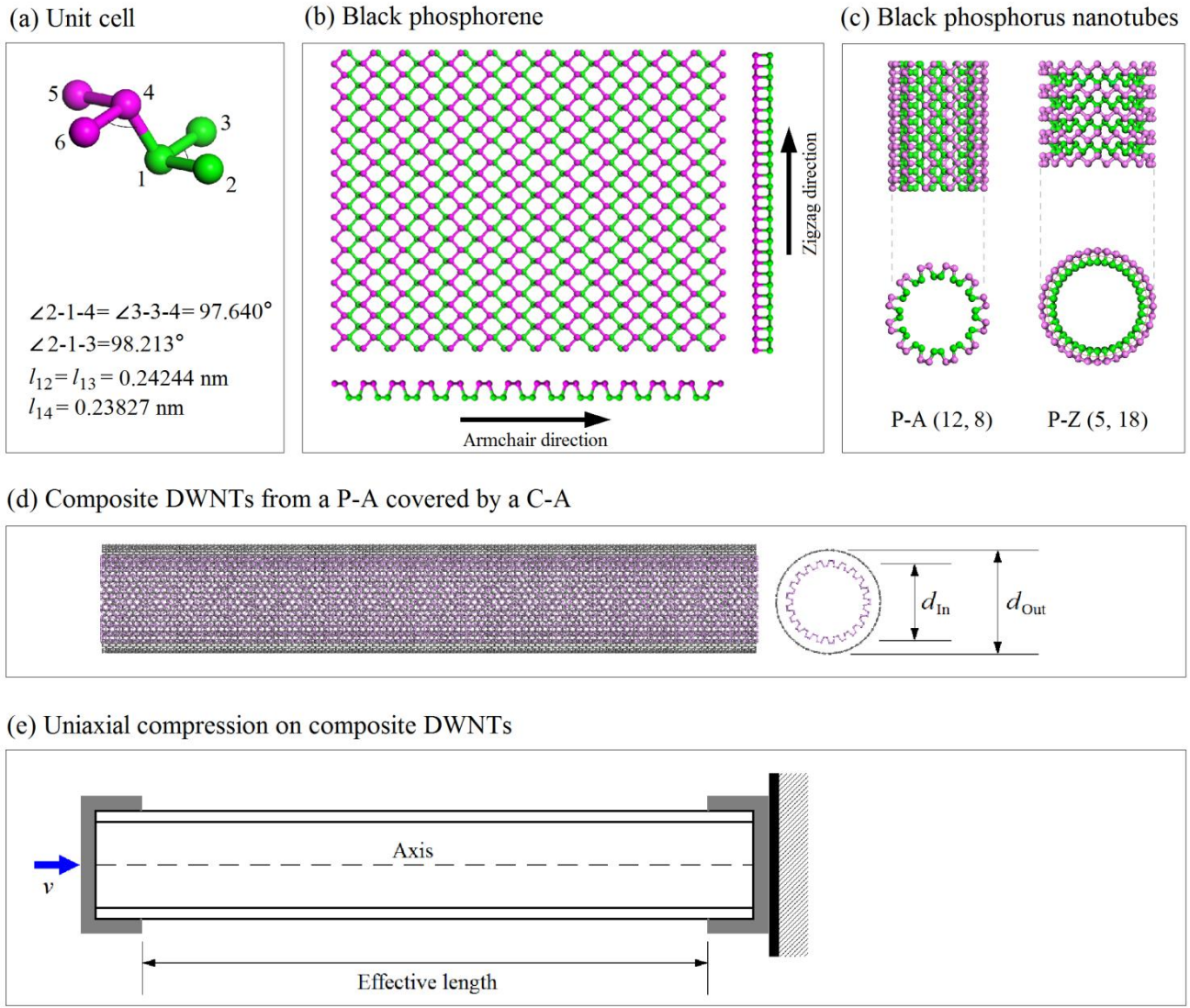

Figure 1 Schematic of composite DWNTs under uniaxial loading. (a) 6-atom Unit cell of black phosphorus; (b) single-layer black phosphorene; (c) Two types of nanotubes from curling black phosphorene along different directions. The chirality of a BPNT is represented with (NA, NZ), the number of unit cells along armchair and zigzag directions, respectively; (d) A composite DWNTs formed by a phosphorus nanotube and a carbon nanotube; (e) Schematic of uniaxial loading of DWNTs.

Table 1. Geometric parameters of the DWNTs in five schemes, in which the ratio between effective length and diameter of BPNT keeps as a constant, i.e., $\alpha$=8. (Length unit: nm)

| Scheme No. | Position | Tube type | Chirality | Length | Effective length | Diameter |
|---|---|---|---|---|---|---|
| S-1 | Inner tube | P-A | (21, 0) | 23.86 | 23.45 | 2.93 |
|  | Outer tube | C-A | (29, 29) | 23.73 | 23.45 | 3.93 |
| S-2 | Inner tube | P-A | (21, 0) | 23.86 | 23.45 | 2.93 |
|  | Outer tube | C-Z | (50, 0) | 23.71 | 23.45 | 3.91 |
| S-3 | Inner tube | C-A | (14, 14) | 23.73 | 23.45 | 1.90 |
|  | Outer tube | P-A | (21, 0) | 23.86 | 23.45 | 2.93 |
| S-4 | Inner tube | C-Z | (25, 0) | 23.79 | 23.45 | 1.96 |
|  | Outer tube | P-A | (21, 0) | 23.86 | 23.45 | 2.93 |
| S-5 | Inner tube | P-Z | (0, 52) | 44.35 | 43.88 | 5.49 |
| | Outer tube | C-Z | (83, 0) | 44.16 | 43.88 | 6.50 |

2.2 Methodology

Three potential functions are involved in description of the interaction among atoms. To show the interaction among the carbon atoms in CNT, AIREBO potential[43] is adopted. The new parameters in Stillinger-Weber (S-W)[44] potential developed by Jiang[45] is used to obtain the atoms interaction within BPNT. The interaction between the carbon atoms and phosphorus atoms on two tubes are described by Lennard-Jones (L-J) potential.[46] For finding the dynamic response of DWNTs under compression, 6 steps of operation are contained, i.e.,

(1) Build DWNTs from a CNT and a BPNT with specified geometry;
(2) Reshape the nanostructure through potential energy minimization (the steepest descending algorithm is adopted);
(3) Fix the degree of freedoms (DOFs) of the atoms nearby the two ends of DWNTs to control the effective length of DWNTs to be a given value;
(4) Put the free atoms (both carbon and phosphorous atoms) in a canonical NVT ensemble with T, e.g., 300K;
(5) After 50 ps of relaxation, move the left end of DWNTs to right with a constant velocity, i.e., in each load step, 0.001nm of movement within one time step followed with 2ps of relaxation;
(6) Record the data of DWNTs during loading, e.g., axial stresses of two tubes, potential energy due to deformation of system, till collapse of the nanosystem.

The time step is 1.0 fs in the present study. Open source code LAMMPS[47] is adopted to fulfill the simulations. The crosssection area of tube is calcuated by A = D×D, where D is the diameter of tube. The volumes of carbon/phosphrous atoms are computed by dividing equally the volume of shell to each atom, the thickness of carbon/phosphrous layer is taken as 0.34nm/0.48nm in mechanical calculation.

## 3 Results and discussion
3.1 Temperature influence

In composite DWNTs from carbon and phosphorus, the two tubes have different thermal expensive coefficients. As the two ends of the composite DWNTs are fixed together at a temperature after a period of relaxation, e.g., 50ps, the internal axial stresses on two tubes appear if the environmental temperature is different from the assembly temperature (the temperature at which the two tubes' ends are fixed (Figure 1e)). Before studying the other mechanical properties or electronic properties of the composite DWNTs, the temperature effect is firstly discussed. In this part, the model for DWNTs is chosen from Table 1, i.e., S-1 with armchair BPNT as inner tube and armchair CNT as outer tube. The environmental temperature is set to be 8, 150, 300 and 400K, respectively.

After 50ps of relaxation at 300K but before loading, the two ends of each tube are fixed. From Figure 2, one can find that the two tubes have non-zero initial stress along axes. For example, in Figure 2a, the initial stresses of BPNT are 0.005 (at 8K), 0.072 (at 150K), 0.140 (at 300K) and 0.178 GPa (at 400K), respectively. It is necessary to demonstrate that the positive value of a compressive stress means that the tube is actually under compression, rather than under tension. For the CNT, the initial stresses (in Figure 2c) are -3.609 (at 8K), -3.804 (at 150K), -3.836 (at 300K) and -3.828 GPa (at 400K), respectively. Hence, CNT is under tension. Especially, at 300K, the two tubes have non-zero initial stress, which implies that only 50ps of relaxation on the system cannot remove the assembly stress. It also says that BPNT has no enough extension whilst CNT does not have enough shrink after

50ps of relaxation.

For BPNT, the compressive stress increases monotonically before rapid drop. The maximal stress of BPNT at 8K is 1.614 GPa after 0.964nm of compressed length along axis. When temperature is higher, the peak value of axial stress becomes lower. For example, at 150K, the peak value of compressive stress on BPNT is 1.457 GPa. If temperature is no less than 300K, the peak value is no more than 1.35 GPa. The reason is that the thermal vibration of the atoms on shells leads to the buckling of shells. During compression, the axial stress of CNT deceases to zero when the compressed length reaches 0.38nm (or 1.62% of engineering strain). As compression keeps going, the tube is under compression as well. The peak value of axial stress of CNT can be higher than 5GPa (Figure 2c), which is higher greater than that BPNT at the same temperature. Similarly, the peak value of axial stress of CNT is lower if at higher temperature.

From Figure 2a and Figure 2c, the two peak values (of BPNT and CNT at the same temperature) appear closely, e.g., no more than 0.002nm of compression difference (or two load steps). From the results we conclude that one tube buckles firstly, and the buckling of the other tube is triggered by the deformation of the buckled tube (see Figure 3). The reason can be found from the geometric parameters of two tubes. For two tubes, they have the same effective length. The radius difference of two tubes equals the equilibrium distance between graphene sheet and black phosphorene sheet, i.e., 0.38332nm. [48] As one tube, e.g., the CNT, buckles, the radial deformation of the tube forces the other tube to have similar deformation. Hence, the composite DWNTs buckling state can be judged from that of either inner tube or outer tube.

In Figure 2b and Figure 2d, the histories of variation of potential energy (VPE) of two tubes under compression are illustrated. For BPNT, the VPE increases monotonically before reaching the maximal value when the temperature is 8K. If at higher temperature, the VPE drops firstly and then increases as load keeps going. It is known that the potential energy of deformation is never negative. Hence, the drop of VPE is caused by further relaxation of the BPNT and interaction between two tubes during loading. Except at 8K, the peak value of VPE of BPNT always appears later than that of axial stress. This is actually due to the local damage of the tube. For example, as the axial stress approaches the peak value, the BPNT has no capability to provide higher stress at higher compressed length. Hence, the drop of stress is due to broken of P-P bonds nearby the location with the most critical deformation (Figure 3). When the number of broken bonds is low, the VPE increases due to increase of surface potential energy. However, if the number becomes high, i.e., the tube is broken heavily, the other area nearby breakage will be relaxed, which leads to sharp drop of deformation potential energy. Hence, the peak value of VPE approaches only after serious breakage of tube. From Figure 2d, the CNT also has serious local deformation after the tube has maximal stress. After jump, VPE may increases in some cases, e.g., BPNT at 8K or CNT at any temperature. It demonstrated that the deformation potential and surface potential increases further after large compression.

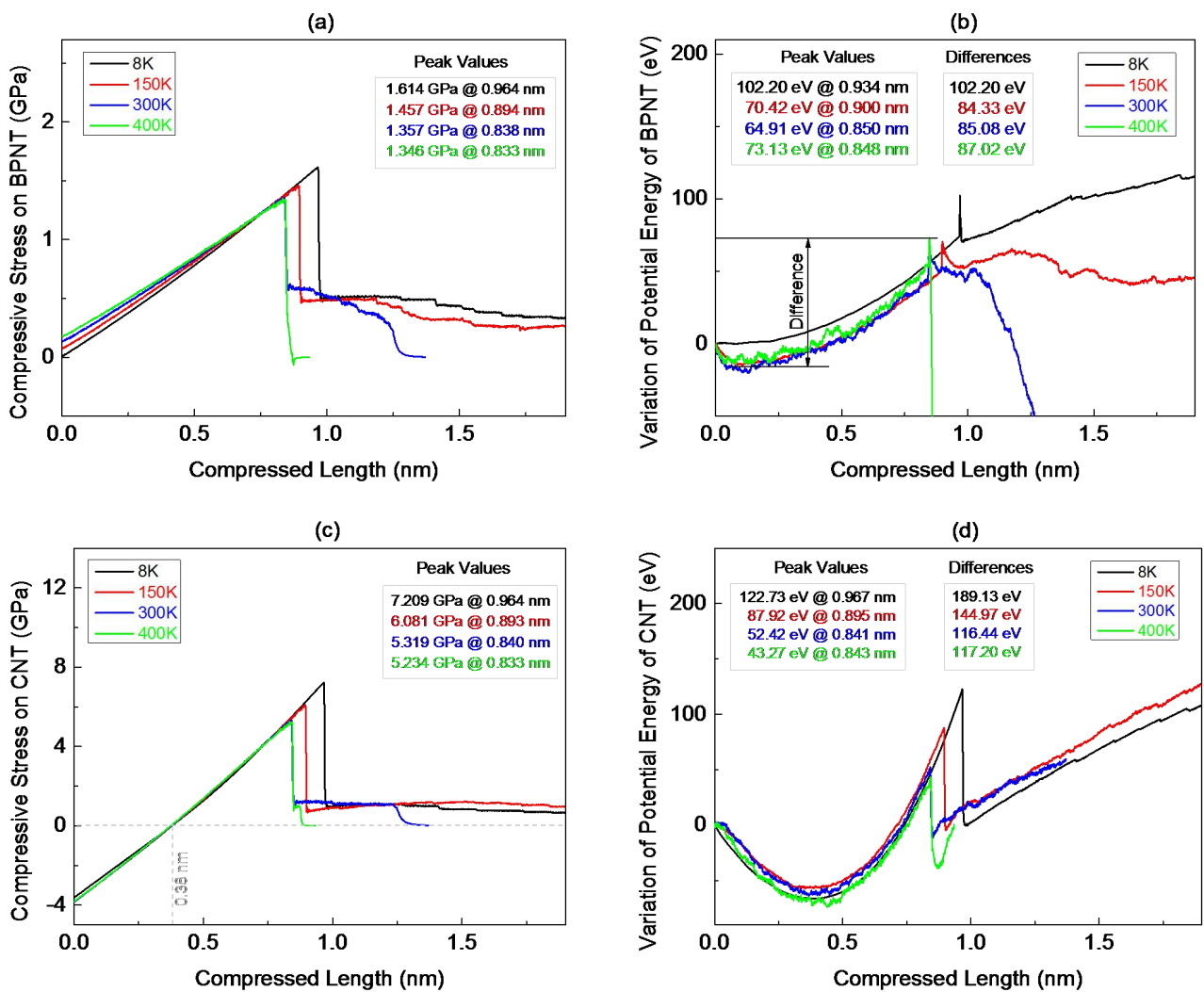

Figure 2  The histories of the axial stress and variation of potential energy of two tubes under compression at different temperature. Difference represents the increment of VPE from minimal to maximal values before jump.

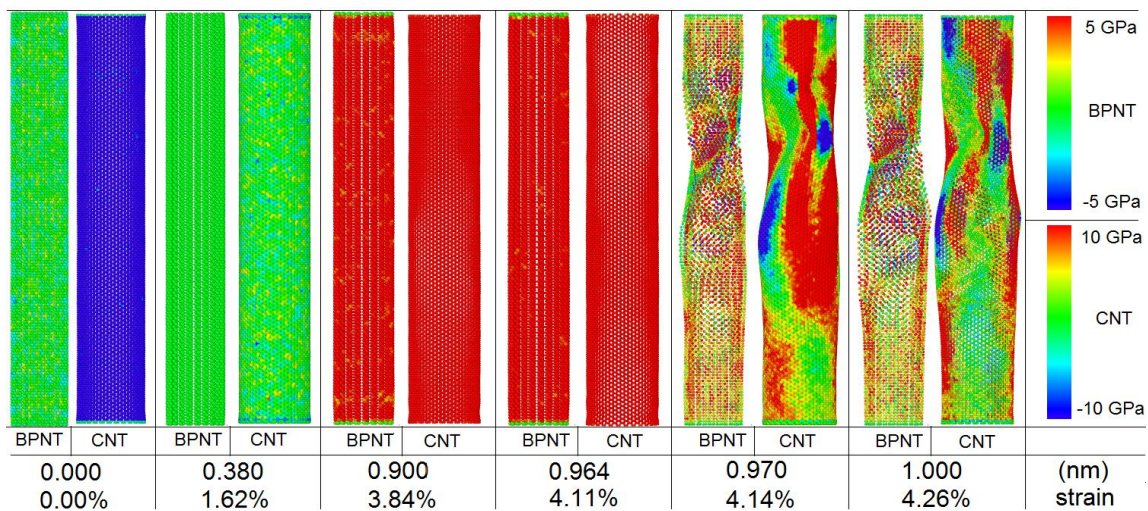

Figure 3  The configurations and stress distributions of composite DWNTs at 8K under compression. Two tubes in S-1 are shown, separately. When engineering strain is over 4.11%, CNT buckles and BPNT is broken.

## 3.2 Slenderness effect

Slenderness, which represents the compliance of a column bearing compression, is essential for the critical buckling deformation of column. To illustrate the influence of slenderness on the buckling state of a tube, here we discussed the influence using the variation of the ratio between the effective length and diameter of BPNT. The ratio is set to be 6, 8, 10 and 12, respectively. The related data are shown in Figure 4.

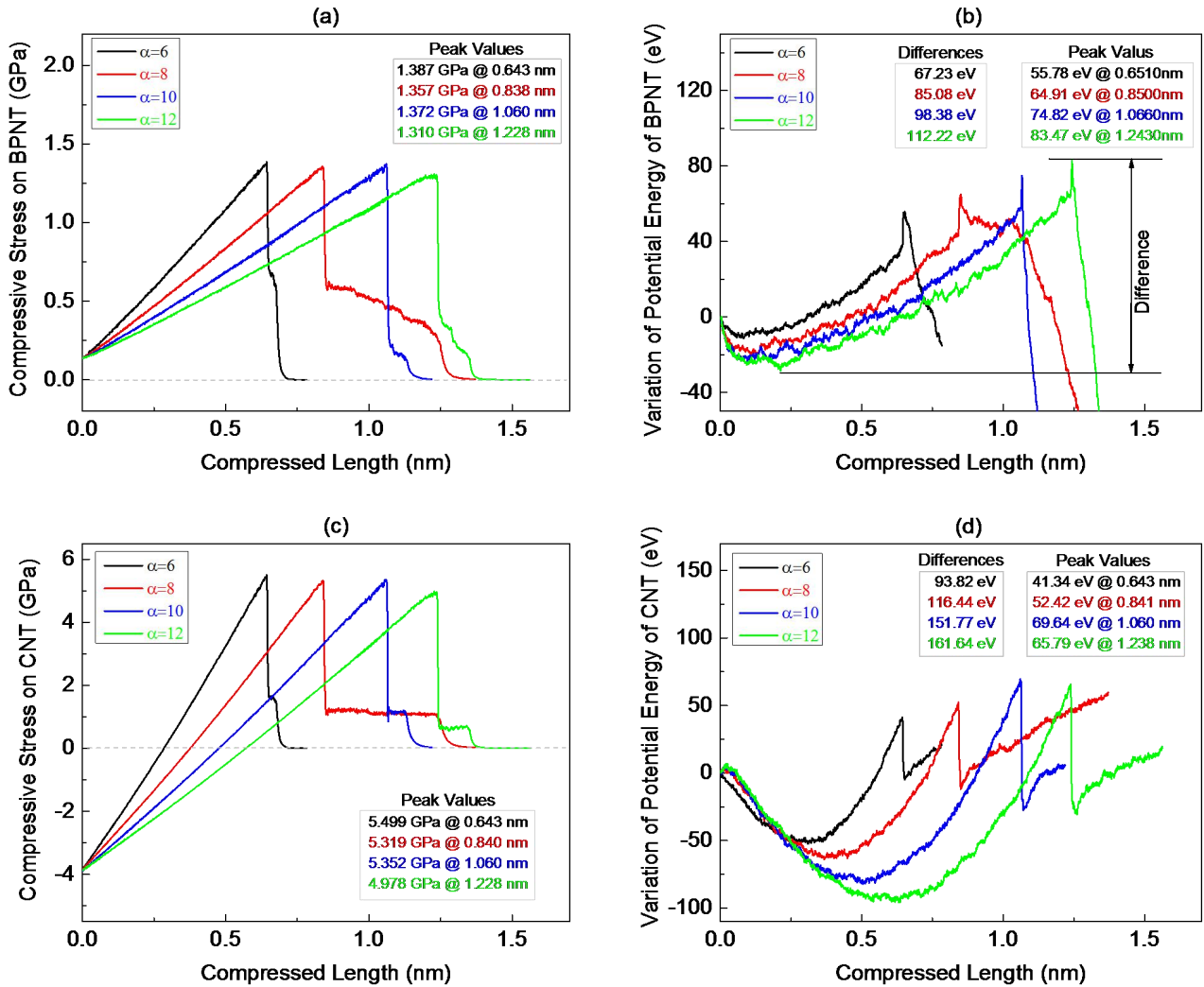

Figure 4 The histories of the axial stress and variation of potential energy of two tubes with different length-diameter ratios under compression.

Figure 4a illustrates the history of compressive stress on the BPNT along axis during loading. The initial values of stresses of tubes with different lengths are the same. The peak values of axial stress on BPNT are different slightly. This is obviously different from that of a single-wall BPNT under compression. [42] For example, the maximal stress on the single-wall BPNT with $\alpha$=12 is the smallest among the four cases ($\alpha$=6, 8, 10 and 12), and the value of peak stress is 1.531 GPa, which is higher than 1.387 GPa, i.e., the maximal stress of the present BPNT in composite DWNTs with $\alpha$=6. As considering the strain of tube with respect to the peak stress as the critical strain, the values of critical strains have slight difference too. For example, for the BPNTs the critical strains are 0.643/17.59 (≈3.66% as $\alpha$=6), 0.838/23.45 (≈3.57% as $\alpha$=8), 1.060/29.31 (≈3.62% as $\alpha$=10) and 1.228/35.17 (≈3.49% as $\alpha$=12), respectively. The critical strains are greater than that of the single-wall BPNT with $\alpha$=12. Hence, tube length has slight influence on strength and stability of BPNT in composite DWNTs with 6<$\alpha$<12. From Figure 4c, the compressive stress of CNT starts from negative to positive during loading. The differences among the peak values of compressive stress on CNT are also small. The critical strains of CNT with different lengths are also different slightly. Two aspects of the curves shown in Figure 4a and Figure 4c are necessary to be demonstrated. One is that the initial stresses of either CNTs or BPNTs are identical, i.e., which does not depend on the tube length. The reason is that both of the relaxation style and relaxation duration (50ps) in the schemes are the same. During relaxation, except two ends of tube, the geometry of the tube varies uniformly. Hence, the initial stress is not sensitive to the length of tube, which is fixed at two ends after such relaxation. The other aspect is that the critical buckling state of the composite DWNTs has weak sensitivity to the length of tube. This character is obviously different from a single tube, e.g., BPNT, which is shown in the work by Cai, Wan, et al.[42]

Figure 4b gives the variation history of potential energy of BPNT during loading. Before approaching the peak value, VPE decreases slowly and then increases monotonically to the peak value. The differences of VPE of BPNTs with different lengths are listed in the figure. It can be found that the differences are approximately proportional to the lengths of tubes. Figure 4d shows that the difference of VPE of CNT is higher for a longer tube. Meanwhile, the VPE curves of CNTs are similar. The critical strains of CNTs are also similar as comparing the data inserted in this figure. A conclusion can be drawn that the deformation potential energy of the tubes in the composite DWNTs depends on the difference of VPE rather than on the peak value of VPE.

3.3 Effect of configurations of composite DWNTs

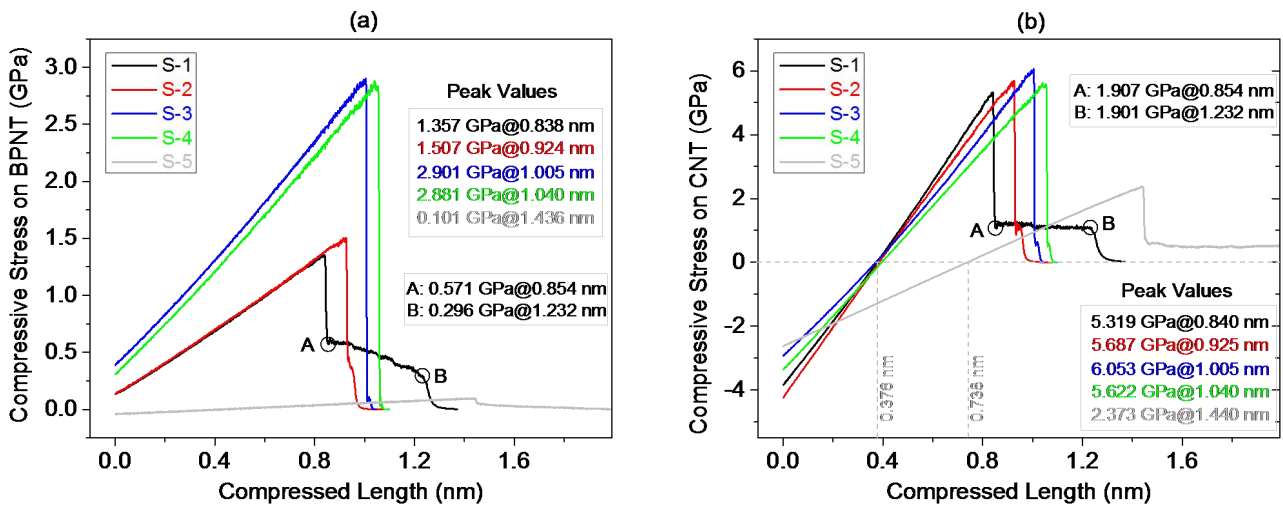

Figure 5 Axial compressive stress on tubes, i.e., (a) BPNT and (b) CNT, during compression with respect to five schemes.

To reveal the effect of configurations of composite DWNTs on its strength and stability, five schemes are considered and the related parameters are listed in Table 1. For the five composite tubes, the ratios between the effective lengths and diameters of BPNTs are identical, i.e., $\alpha$=8. In the first four models, the same BPNT (P-A) is adopted, hence the effective lengths are the same (23.45 nm). For the fifth model, P-Z is used. Due to higher diameter of a stable P-Z at 300K, the effective length is 43.88 nm.

In S-1 and S-2, BPNT (P-A) acts as an inner tube and covered by C-A and C-Z, respectively. From Figure 5a, we find that the peak value of axial compressive stress on BPNT in S-1 is lower than that in S-2. Therefore, one can conclude that the difference of peak values of BPNT in the two schemes is due to the difference of configurations of the outer tube, i.e., C-A and C-Z. Before compressed length approaching 0.838nm, the axial stress curves for BPNT in S-1 and S-2 are identical. It implies that the two CNTs have the same axial modulus under small deformation. After buckling, the BPNT is damaged and the axial stresses of CNT and BPNT in S-1 jump down quickly. After jump, the axial stresses are not decreasing quickly. For example, on the black curves in Figure 5, from point A to point B, the stresses on CNT changes slower than that on BPNT. It is because the breakage of BPNT prevents the postbuckling [49,50] of CNT during the period (Figure 6a).

In S-3 and S-4, the same BPNT is supported by different CNTs (C-A and C-Z) on the inner surface, respectively. From the blue and green curves in Figure 5a, one can find that the peak values of BPNT supported by different CNTs have slight difference. However, the two curves do not overlap, which implies that the BPNT supported by C-A will have a higher modulus and higher axial stress than supported by C-Z when they have the same axial compression. The reason is that the two CNTs have different axial moduli due to low diameters.

As comparing S-1 and S-3 (or S-2 and S-4), the difference between the maximal axial stresses of BPNT is great, e.g., the peak value of stress in S-3 is more than twice of that in S-1. Similar great difference also exists between S-4 and S-2. But, the critical axial strain (or compressed length) of BPNT in S-3 is only ~20% higher than that in S-1. It demonstrates that the BPNT acting as an outer tube will have a higher modulus and higher critical axial strain than that as an inner tube when it is supported by a CNT. The major reason is that BPNT has higher freedom of deformation along radial direction when its outer surface is not restricted whilst the deformation of the inner surfaces is confined. From our previous work,[41] the collapse of P-A is due to both of higher extension of P-P bonds on the outer surface and strong attraction among the phosphorus atoms on the inner surface. As the inner surface is confined by CNT, the bond lengths and bond angles on the inner surfaces have lower variation. Hence, the critical compressive stress on BPNT along axis is enhanced. It is necessary to mention that in S-3 (or in S-4), BPNT acts as an outer tube, which implies that the slenderness of the composite DWNTs in S-3 is actually lower than that of DWNTs in S-1. Therefore, BPNT (P-A) supported on the inner surface will have higher mechanical properties (both stiffness and strength).

In the fifth model, BPNT (P-Z) is protected by a CNT (C-Z) on the outer surface. Due to the diameter of P-Z being about 1.8 times of that of P-A in the first four models (Table 1), the ratio between the thickness and diameter of composite tube is lower than that in the first four models. Hence, the composite tube in S-5 has lower slenderness than any one of the first four models. From Figure 5, we find that the maximal compressed length of P-Z is 1.44nm, or 3.28% of critical engineering strain of the BPNT. The critical strain is actually close to that of BPNT in S-1(0.838/23.45=3.57%), S-2(0.924/23.45=3.94%), S-3(1.005/23.45=4.29%) or S-4(1.040/23.45=4.43%). From our previous study, the BPNT has low stability at 300K if it is neither fixed nor loaded. But currently, the compressive strain of P-Z can be so high. It implies that the P-Z can also be used as a component in NEMS when it is protected by a CNT on its outer surface.

Figure 5b shows the axial stress histories of CNTs in different models. From the maximal compressed lengths of CNTs, we know that the coupling deformation of CNT and BPNT is still essential to the buckling and strength of the composite DWNTs with any type of configurations. Some representative snapshots of the composite DWNTs in S-1, S-3 and S-5 are shown in Figure 6. It can be found that the BPNT will break after buckling. If BPNT is an outer tube in the composite DWNTs, some phosphorus atoms may leave away from the tube after serious breakage (see the red circles in Figure 6b). This phenomenon should be avoided in NEMS. Hence, we suggest such composite DWNTs as shown in S-1, S-2 or S-5 (see Figure 6a, c).

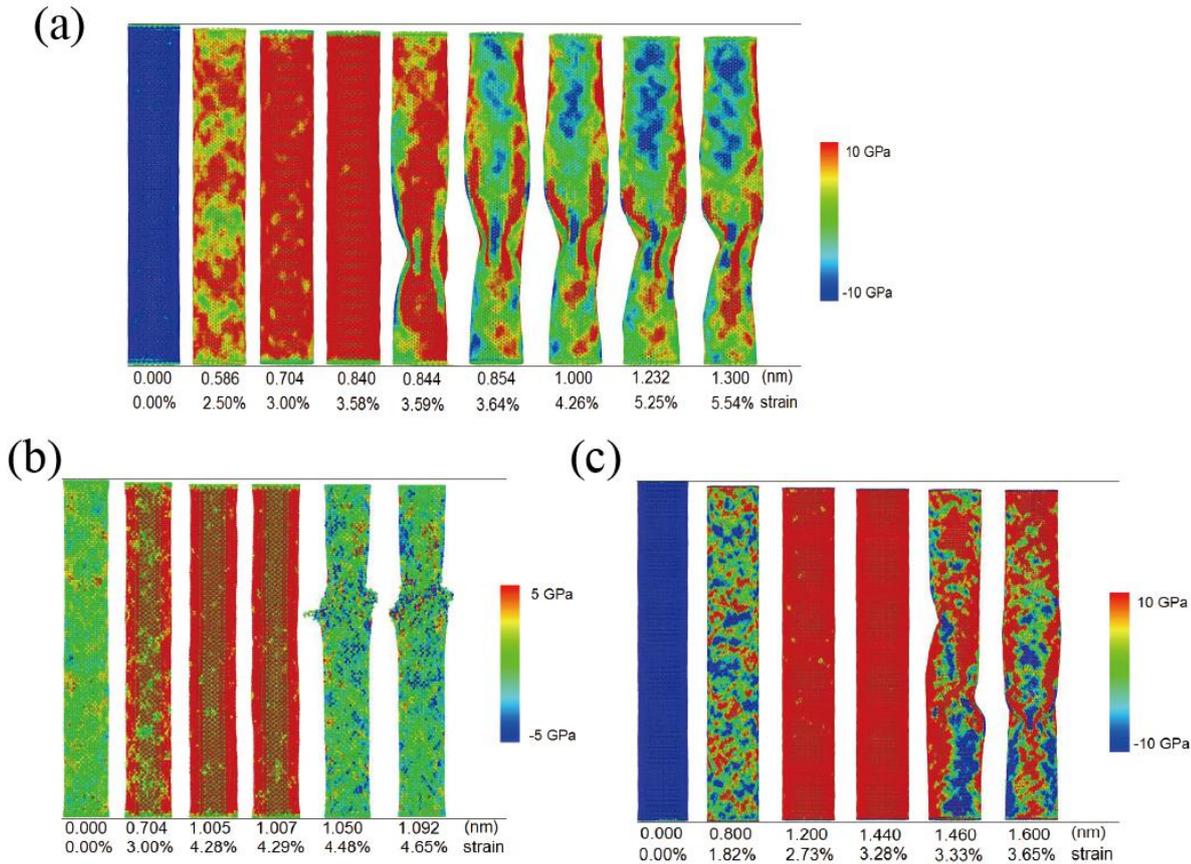

Figure 6 The configurations and stress distributions of outer tubes in composite DWNTs during compression. (a-c) are composite DWNT structures corresponding to S-1, S-3, and S-6 , respectively.

**4 Conclusions**

Due to weak P-P interaction in black phosphorus nanotube, we use carbon nanotube to protect on the BPNT. From the numerical results on considering the influence of temperature, slenderness and configurations of the composite DWNTs from CNT and BPNT, some conclusions are drawn, i.e.,

1) The two tubes (CNT and BPNT) have non-initial stresses (assembly stress) as they are not relaxed fully before loading. The maximal axial compressive stresses of two tubes are lower at higher temperature;
2) After peak value of axial stress, one tube buckles firstly, and the other is actuated to buckle immediately;
3) The maximal value of VPE appears later than that of axial stress on BPNT because breakage of few P-P bonds leads to increase of VPE but decrease of axial stress. VPE approaches maximum only after serious breakage of BPNT;
4) Tube length has slight influence on strength and stability of the BPNT in composite DWNTs with $6<\alpha<12$. This is different from that of a single-wall BPNT under compression;
5) BPNT as an outer tube in the composite DWNTs has higher modulus and higher critical axial strain than those of it as an inner tube. However, serious breakage of the BPNT leads to collapse of the nanosystem, and some phosphorus atoms may leave away from the DWNTs. It should be avoided in NEMS. Hence, BPNT as inner tube in the composite DWNTs is recommended.

Acknowledgement

Financial support from the National Natural Science Foundation of China (Grant No. 11372100, 11502217), the China Postdoctoral Science Foundation (No. 2015M570854, 2016T90949) and the Australian Research Council (Grant No. DP140103137) is greatly acknowledged.